\documentstyle[amstex,amssymb,epsf,aps,multicol,prb]{revtex}

\def\bm{\begin{multicols}{2}}  \def\em{\end{multicols}}

\input epsf

\begin{document}
\title{Theory of superconductor with $\kappa $ close to $1/\sqrt{2}$}
\author{I. Luk'yanchuk}
\address{L.D.Landau Institute for Theoretical Physics, Moscow, Russia}
\address{Institut f\"{u}r Theoretische Physik, RWTH-Aachen, Templergraben 55, 52056\\
Aachen, Germany}
\author{(\today)}
\maketitle

\begin{abstract}
As was firstly shown by E. Bogomolny, the critical Ginzburg-Landay (GL)
parameter $\kappa =1/\sqrt{2}$ at which a superconductor changes its
behavior from type-I to type-II, is the special highly degenerate point
where Abrikosov vortices do not interact and therefore all vortex states
have the same energy. Developing a secular perturbation theory we studied
how this degeneracy is lifted when $\kappa $ is slightly different from $1/%
\sqrt{2}$ or when the GL theory is extended to the higher in $T-T_{c}$
terms. We constructed a simple secular functional, that depends only on few
experimentally measurable phenomenological parameters and therefore is quite
efficient to study the vortex state of superconductor in this transitional
region of $\kappa $. Basing on this, we calculated such vortex state
properties as: critical fields, energy of the normal-superconductor
interface, energy of the vortex lattice, vortex interaction energy etc. and
compared them with previous results that were based on bulky solutions of GL
equations.
\end{abstract}

\bm

\section{Introduction}

Although the Ginzburg-Landau (GL) theory covers all the variety of
superconductors, both of the type I with GL parameter $\kappa <1/\sqrt{2}$
and of the type II with $\kappa >1/\sqrt{2}$, the most theoretical studies
of the vortex state deal with the case of $\kappa \gg 1/\sqrt{2}$ since at $%
\kappa \gg 1/\sqrt{2}$ the GL equations are simplified substantially and
also the grand majority of type-II superconducting materials, including
High-Tc superconductors, correspond to this limit.

Studies of superconductors in the transitional region of small $\kappa \sim
1/\sqrt{2}$ done mostly in the 70-s were based on the solution of the full
system of GL equations and require bulky calculations. Meanwhile, E.
Bogomolny proposed in 1976 an elegant way to operate with similar problem of
the string theory \cite{Bog} and showed that at the special point $\kappa =1/%
\sqrt{2}$ the order of the equations can be reduced and all the vortex
states with arbitrary located vortices have the same energy when the applied
field is equal to the critical field $H_{c}$.

Historically the Bogomolny approach was done for the high-energy physics and
the superconducting community was unaware of it even when L. Jacobs and C.
Rebbi \cite{Jac} reformulated the Bogomolny equations in terms of
superconductivity and demonstrated that they can be written in a form of
nonlinear electrostatic equations of the Boltzman plasma that we are calling
as Bogomolny Jacobs and Rebbi (BJR) equation. Only very recently the
Bogomolny method was used to study vortices in mesoscopic disks \cite{Akker}%
\ and to calculate the structure of multi-quanta vortices \cite{Efanov} in
superconductors with $\kappa =1/\sqrt{2}$.

In the present paper we derive a regular way to treat the vortex state of
superconductors with $\kappa $ close to $1/\sqrt{2}$ basing on the BJR
approach. Our basic idea is to consider the highly degenerate Bogomolny
state at $\kappa =1/\sqrt{2}$, $H=H_{c}$ as zero approximation and then to
account for deviation from this point via the secular perturbation method
that lifts the degeneracy and selects the most stable vortex configuration.
In Sec. \ref{Sect:Bog} we give a brief overview of the Bogomolny method and
then, in Sec. \ref{Sect:VortLat}, present the solutions of BJR equation for
the different vortex lattices. In Sec. \ref{Sect:Perturbation} we develop
the secular perturbation approach. The following perturbations can lift the
degeneracy.

(i) Deviation of $\kappa $ from $1/\sqrt{2}$ that is accounted by a small
parameter. 
\begin{equation}
\gamma =\kappa ^{2}-\frac{1}{2}\approx \sqrt{2}\kappa -1.
\end{equation}

(ii) Deviation of the applied field from $H_{c}$.

(iii) Next in 
\begin{equation}
t=T/T_{c}-1
\end{equation}
corrections to the GL functional.

(iy) Thermal fluctuation effects.

(y) Finite size and demagnetization effects.

We consider only the first three contributions and show that they can be
incorporated in a very simple secular functional that acts on the Bogomolny
degenerate solutions and looks like a six-order polynomial for the amplitude
of the order parameter. This functional depends only\ on few
phenomenological parameters that can be found from experiment and that
completely determine the behavior of superconductors with $\kappa \sim 1/%
\sqrt{2}$ in a magnetic field.

Certain properties of superconductors with $\kappa \sim 1/\sqrt{2}$ were
calculated either from the GL theory extended to low temperatures or from
the microscopic Gorkov equations. These calculations, overviewed in Sec. \ref
{Sect:Previous}, were however dealing either with cumbersome analytical
expansions or with numerical computations that both are difficult to catch
on. It is therefore of interest to recalculate these properties in a
systematic perturbation way and compare them with the older results.

In Sec. \ref{Sect:VortEnergy} we calculate the following parameters of a
superconductor with $\kappa \sim 1/\sqrt{2}$:

a) Critical fields $H_{c1}$, $H_{c2}$ and $H_{c}$.

b) Energy of the NS interface.

c) Energy of the regular vortex lattice as function of the applied field.

d) Energy of the N-quanta vortex.

e) Vortex interaction that can have unconventional attractive character.

Basing on these calculations we discuss the possible scenarios of the
normal-superconducting (N-S) transition in a magnetic field for a
superconductor with $\kappa \sim 1/\sqrt{2}$ (Sec.\ref{Sect:Discussion})
that occurs either directly (like in a type-I superconductor) or via
formation of the intermediate vortex (V) state (like in a type-II
superconductor). The actual scenario depends on the relative strength and
sign of coefficients in the perturbation functional that can be extracted
from experiment. We calculate the location of the triple point $L$ on the $%
H-T$ plane where the direct N-S transition splits into N-V and V-S
transitions and a superconductor changes its behavior from type-I to
type-II. The important feature is that, the N-V and V-S transitions close to
point $L$ can be both continuous and discontinuous unlike the traditional
type-II superconductor with $\kappa \gg 1/\sqrt{2}$. We calculated the
location of tricritical points $T_{2}$ and $T_{1}$ where the N-V and V-S
transitions change their character from continuous to discontinuous.

\section{Previous study}

\label{Sect:Previous}

\subsection{Theory}

Already in his pioneering work \cite{Abr} Abrikosov noted that solution of
the GL equations at $\kappa \sim 1/\sqrt{2}$\ is a separate and quite
complicate problem. Since that various related theoretical investigations
that are partially reviewed in \cite{Hubner,BrandtRev,Saint} were done. The
first series of investigations dealt with an expansion of the BCS free
energy close to $H_{c2}$ over a small parameter $H-H_{c2}$. The magnetic and
thermodynamic properties of superconductor close to $H_{c2}$ were calculated
for dirty \cite{Maki,DeGennes} and intrinsic \cite{MakiIntr}
superconductors. The most complete calculations of this type are given in 
\cite{Eilenberger}. The possibility to have a discontinuous N-V transition
in a superconductor with $\kappa \sim 1/\sqrt{2}$ was firstly indicated in 
\cite{Maki}.

On the basis of BCS theory Tewordt and Neumann calculated the
low-temperature corrections to GL functional \cite{TewGen,TewPR,TewHc1} and
found the upper \cite{TewHc2} and low \cite{TewHc1} critical fields with an
accuracy $t^{2}$ at arbitrary $\kappa $.

Basing on this extended GL functional, A. E. Jacobs \cite{Jac123} considered
a superconductor with $\kappa \sim 1/\sqrt{2}$ and calculated the NS
interface energy, the energy of single and double quantized vortex. He
obtained that at certain conditions the vortices in a type-II superconductor
attract each other and predicted the discontinuity of the V-S and N-V
transitions. The analogous result was also obtained by Hubert \cite{Hubert}%
\thinspace .

Gro\ss mann and Wissel \cite{Gross} calculated the free energy of a
superconductor with $\kappa \sim 1/\sqrt{2}$ close to $H_{c2}$ using the
extended (although not complete) functional of Tewordt and Neumann. They
founded a discontinuity of V-S transition in a limit of the dense vortex
lattice. All the above conclusions were reproduced by Brandt \cite
{BrandtLatAll} who developed a variational numerical method to solve the
Gorkov's equation for vortex lattices for all possible values of $H$, $T$
and $\kappa $.

Recently Ovchinnikov \cite{Ovch} carefully derived the coefficients of the
extended GL functional from a microscopic theory for different types of the
electron scattering. He considered an expansion of the free energy near $%
H_{c2}$ up to the order of $(H-H_{c2})^{3}$ and specified the case when at $%
\kappa \sim 1/\sqrt{2}$ the N-V transition has a discontinuous character.

In the present article we reproduce the above results in a more simple way,
basing on the Bogomolny treatment of superconductors with $\kappa =1/\sqrt{2}
$.

\subsection{Experiment}

The superconducting metals Ta, Nb, In and Pb with $\kappa $ close to $1/%
\sqrt{2}$ were intensively studied in the 60-s and 70-s. The variation of $%
\kappa $ was achieved either by dissolving of foreign atoms of N, Tl, Bi or
by preparation of samples with different defect concentration. We refer to
magnetic \cite{NbMagn} calorimetric \cite{NbCal} and neutron diffraction 
\cite{NbNeutr,NbNeutr2} experiments in pure Nb ($\kappa \sim 0.85-0.96$), to
magnetic measurements in TaN ($\kappa \sim 0.35-1.53$) \cite{(Ta-Nb)Nmagn},
in Nb ($\kappa \sim 0.78-1.03$) \cite{(Ta-Nb)Nmagn} and in InBi ($\kappa
\sim 0.76-1.46$) \cite{InBiMagn} and to direct observation of vortices in
PbTl ($\kappa \sim 0.43-1.04$) \cite{PbTlDecor} and in PbIn ($\kappa \sim
0.76-1.46$) \cite{PbInDecorFirst,PbInDecor} by decoration. References to
other related experiments can be fond in \cite{Hubner,BrandtRev}.

The fact that the V-S transition can be of the first order at $\kappa \sim 1/%
\sqrt{2}$ was discovered already in the early magnetic and thermodynamic
experiments \cite{NbMagn,NbCal,InBiMagn}. The detailed magnetic study of a
superconductor that changes its behavior from type I to type II was done for
tantalum samples with some amount of dissolved nitrogen \cite{(Ta-Nb)Nmagn}.
A discontinuity of the vortex lattice parameter at the V-S transition was
observed in neutron-scattering experiments \cite{NbNeutr,NbNeutr2}.

The convincing confirmation of discontinuity of the V-S transition in
superconductors with $\kappa \sim 1/\sqrt{2}$ was done by a direct
observation the vortex domains inside the Meissner phase \cite
{PbTlDecor,PbInDecorFirst,PbInDecor}. Such coexistence of different phases
is known to be a signature of the first order transition between them. This
intermediate-mixed domain structure was interpreted in \cite{PbTlDecor} in
terms of a long-range vortex attraction.

The discontinuity of the V-S transition provided by an attractive
interaction between vortices is therefore a well established fact. Meanwhile
the ground state of the vortex lattice and the configuration of domains of
the mixed-intermediate phase are still unclear. Although the decoration
experiments \cite{Hubner,BrandtRev,PbTlDecor,PbInDecorFirst,PbInDecor}
demonstrate the very peculiar magnetic textures, including vortex
segregation and clustering into lamellar and drop-like domains, no
systematic study of this question that take into account the demagnetization
and finite-size effects was done. We believe that our calculations of the
vortex energy in the bulk superconductor with $\kappa \sim 1/\sqrt{2}$ can
be extended to simulation of magnetic textures in the realistic finite-size
samples.

\section{GL functional at $\protect\kappa ^{2}=1/2$ and BJR equation}

\label{Sect:Bog}

In this Section we describe the Bogomolny procedure \cite{Bog} that allows
to simplify the GL functional and to reduce the order of the GL equations at 
$\kappa =1/\sqrt{2}$. Jacobs and Rebbi \cite{Jac} formulated the Bogomolny
equations in a simple form of the nonlinear Poisson equation that we shall
call as BJR equation. We discuss the properties of the vortex solutions of
the BJR equation and their interpretation in terms of electrons in Boltzman
plasma given in \cite{Efanov}.

We start from the conventional GL functional: 
\begin{equation}
{\cal F}=\alpha \left| \Psi \right| ^{2}+\frac{g}{2}\left| \Psi \right|
^{4}+K\left| {\bf D}\Psi \right| ^{2}+\frac{B^{2}}{8\pi }-\frac{BH_{0}}{4\pi 
},  \label{LGL}
\end{equation}
where 
\[
\alpha =\alpha _{1}t,{\bf \quad D}=\nabla -i\frac{2e}{c\hbar }{\bf A,}\quad 
{\bf B}=rot{\bf A.}
\]

Refer first to the characteristic parameters of superconductor. In the
uniform state the superconducting order parameter takes the equilibrium
value: 
\begin{equation}
\Psi _{0}=\left( -\frac{\alpha }{g}\right) ^{1/2}.
\end{equation}
Ratio of the penetration depth and coherence length: 
\begin{equation}
\delta =\left( -\frac{c^{2}\hbar ^{2}}{32\pi Ke^{2}}\frac{g}{\alpha }\right)
^{1/2},\quad \xi =(-\frac{K}{\alpha })^{1/2}
\end{equation}
gives the GL parameter: 
\begin{equation}
\kappa =\frac{\delta }{\xi }=\frac{1}{(32\pi )^{1/2}}\frac{c\hbar g^{1/2}}{%
\left| e\right| K}.\   \label{Param}
\end{equation}
Thermodynamic critical field and the upper critical field are written as: 
\begin{equation}
H_{c}=-\left( \frac{4\pi }{g}\right) ^{1/2}\alpha ,\quad H_{c2}=-\frac{%
c\hbar }{2\left| e\right| D}\alpha =\sqrt{2}\kappa H_{c}.  \label{Hc}
\end{equation}
Note that the commonly used expression for the low critical field: 
\begin{equation}
H_{c1}=\frac{\Phi _{0}}{4\pi \delta ^{2}}\ln \kappa =H_{c}\frac{\ln \kappa }{%
\sqrt{2}\kappa }.
\end{equation}
is valid for $\kappa \gg 1/\sqrt{2}$ and is not applicable in our case.
Corresponding expression for $H_{c1\text{ }}$at $\kappa \sim 1/\sqrt{2}$
will be obtained in Sec. \ref{Sect:VortEnergy}. Introduce now the
dimensionless variables that are slightly different from the commonly used
in the GL theory: 
\begin{eqnarray}
\psi  &=&\frac{\Psi }{\Psi _{0}},\qquad r=\frac{R}{\delta \sqrt{2}}, 
\nonumber \\
b &=&\sqrt{2}\kappa \frac{B}{H_{c}},\qquad h_{0}=\sqrt{2}\kappa \frac{H_{0}}{%
H_{c}},\qquad {\bf a=}\ \frac{\kappa {\bf A}}{\delta H_{c}},  \nonumber \\
f &=&\frac{{\cal F}}{H_{c}^{2}/8\pi }\kappa ^{2}+\kappa ^{2}.  \label{dmv}
\end{eqnarray}
The GL functional (\ref{LGL}) in this variables takes the form: 
\begin{equation}
\ f=\kappa ^{2}(\left| \psi \right| ^{2}-1)^{2}+\left| (\ \nabla -i{\bf a}%
)\psi \right| ^{2}+\ (\frac{b^{2}}{2}-bh_{0}).\ 
\end{equation}
It is convenient to use the complex variables: 
\begin{eqnarray}
\zeta ,\overline{\zeta } &=&x\pm iy,  \label{Var} \\
\partial ,\overline{\partial } &=&\frac{1}{2}(\nabla _{x}\mp i\nabla _{y}), 
\nonumber \\
a,\overline{a} &=&\frac{1}{2\ }(a_{x}\mp ia_{y})  \nonumber
\end{eqnarray}
in which the GL functional is written as: 
\begin{eqnarray}
\ f &=&\ 2\left| (\partial -ia)\psi \right| ^{2}+\ 2\left| (\overline{%
\partial }-i\overline{a})\psi \right| ^{2}  \label{Dimless} \\
&&+\kappa ^{2}(\left| \psi \right| ^{2}-1)^{2}\ +\ (\frac{b^{2}}{2}-bh_{0})\
\   \label{FF}
\end{eqnarray}
To catch the special properties of the GL functional at $\kappa =1/\sqrt{2}$
one can integrate the first term in (\ref{Dimless}) by parts. Using that: 
\begin{equation}
\int f\overline{\partial }gdS=-\int g\overline{\partial }fdS+\frac{i}{2}%
\oint fgd\overline{\zeta }
\end{equation}
and that: 
\begin{equation}
b=-2i(\partial \overline{a}-\overline{\partial }a)  \label{Field}
\end{equation}
one gets the substitution: 
\begin{equation}
2\left| (\partial -ia)\psi \right| ^{2}\rightarrow 2\left| (\overline{%
\partial }-i\overline{a})\psi \right| ^{2}+\left| \psi \right| ^{2}b.
\label{part}
\end{equation}
We neglect the contribution of the surface currents that are important for
finite-size effects considered in \cite{Akker}. Finally, one comes to the
alternative expression for $f$: 
\begin{eqnarray}
f &=&4\left| (\overline{\partial }-i\overline{a})\psi \right| ^{2}+\frac{1}{2%
}\left( b+\left| \psi \right| ^{2}-1\right) ^{2}  \label{Alt} \\
&&+\ \gamma (\left| \psi \right| ^{2}-1)^{2}\ +(1-h_{0})b,  \nonumber
\end{eqnarray}
where $\gamma =\kappa ^{2}-1/2$. When $\gamma =0$ and $h_{0}=1$ (i.e. $%
H=H_{c}$) the functional (\ref{Alt}) reduces to the sum of two square terms.
The absolute minimum is achieved when both these terms are equal to zero
i.e. when the following equations are satisfied: 
\begin{equation}
(\overline{\partial }-i\overline{a})\psi =0  \label{Pot}
\end{equation}
and 
\begin{equation}
1-\psi \overline{\psi }=-2i(\partial \overline{a}-\overline{\partial }a)=b.
\label{Second}
\end{equation}
Substitution of $\overline{a}$ from (\ref{Pot}): 
\begin{equation}
\overline{a}=-i\overline{\partial }\ln \psi   \label{aln}
\end{equation}
to (\ref{Second}) gives the BJR equation: 
\begin{equation}
\frac{1}{2}\nabla ^{2}\ln \left| \psi \right| ^{2}=\left| \psi \right|
^{2}-1+2\pi \sum N_{i}\delta ({\bf r}-{\bf r}_{i}).  \label{Elstatalt}
\end{equation}
Firstly introduced in \cite{Efanov} the $\delta $-function terms correspond
to the $N_{i}$-quanta vortices located at ${\bf r}={\bf r}_{i}$ where $\psi $
gets the phase wind $2\pi N_{i}$ and $\ln \left| \psi \right| $ has a $\ln
r^{N}\,$singularity. As follows from Eq. (\ref{Second}) the distribution of
magnetic field inside a sample is uniquely related with the amplitude of the
order parameter: 
\begin{equation}
b({\bf r})=1-\left| \psi ({\bf r})\right| ^{2}.  \label{MF}
\end{equation}
The BJR equation can be alternatively written as: 
\begin{equation}
\nabla ^{2}\varphi =e^{2\varphi }-1+2\pi \sum N_{i}\delta ({\bf r}-{\bf r}%
_{i}).  \label{Elstat}
\end{equation}
where $\varphi =\ln \left| \psi \right| $. This form has a simple
interpretation \cite{Efanov}. It describes the screening of electrons in a
classical Boltzman plasma that consists of the positive ionic background
with potential $1/4\pi $.

BJR equation (\ref{Elstatalt}) defines the amplitude of the superconducting
order parameter $\left| \psi ({\bf r})\right| $ and the magnetic field $b(%
{\bf r})\ $at $\gamma =0$ $\ $and at $h_{0}=1$ as a function of position of
the vortices ${\bf r}_{1}...{\bf r}_{n}$. All these vortex solutions
correspond to the absolute minimum of the functional (\ref{Alt}) and,
therefore, $\gamma =0$ and $h_{0}=1$ is a special highly degenerate point
where all the vortex states have the same energy.

This infinite degeneracy over ${\bf r}_{1}...{\bf r}_{n}$ is lifted if one
goes either beyond $\gamma =0$, $h_{0}=1$ or beyond the GL approximation.
When $\gamma =0$ and $h_{0}>1$ the absolute minimum of (\ref{Alt}) is the
normal state with $b=h_{0}$ and $\left| \psi \right| =0$. When $\gamma =0$
and $h_{0}<1$ the absolute minimum of (\ref{Alt}) is the uniform
superconducting state with $b=0$ and $\left| \psi \right| =1$. To find the
vortex states when $\gamma \neq 0$ one should account for the term $\gamma
(\left| \psi \right| ^{2}-1)^{2}$ as the secular perturbation that lifts the
degeneracy. This will be done in Sec. \ref{Sect:Perturbation} together with
account of the low temperature corrections to the GL functional.

\section{Vortex state at $\protect\kappa ^{2}=1/2$}

\label{Sect:VortLat}

\subsection{General}

In this Section we discuss a particular class of solutions of BJR equation (%
\ref{Elstatalt}) when $N-$ quanta vortices are packed into the regular
lattice with basis vectors ${\bf a}_{1},{\bf a}_{2}$. We will need these
solutions in Sec. \ref{Sect:VortEnergy} as zero approximation of the
perturbation theory to find the most stable vortex configuration beyond the
Bogomolny point. The unit cell area $S=a_{1}a_{2}\sin \alpha $ ($\alpha =%
{\bf a}_{1}{}^{\wedge }{\bf a}_{2}$) carries the flux $2\pi N$ and therefore
is related with the average induction as: 
\begin{equation}
S=2\pi N/\overline{b}.  \label{S}
\end{equation}

The value of $\overline{b}$ and $S$ varies from $\overline{b}=0$, $S=\infty $
(almost non-overlapping vortices) to $\overline{b}=1$, $S=2\pi N$ (dense
vortex lattice). We consider both the limits analytically. We use a
numerical procedure to treat the case of an arbitrary lattice.

\subsection{One vortex}

The axially-symmetric distribution of the order parameter $g_{N}(r)=\left|
\psi (r)\right| $ inside the $N$-quanta vortex is calculated from the radial
version of BJR equation (\ref{Elstatalt}): 
\begin{equation}
g_{N}^{\prime \prime }=\frac{g_{N}^{\prime 2}}{g_{N}}-\frac{g_{N}^{\prime }}{%
r}+g_{N}^{3}-g_{N}  \label{Rad}
\end{equation}
with the boundary conditions: $g_{N}(0)=0$, $g_{N}(\infty )=1$ and
asymptotes: 
\begin{eqnarray}
g_{N} &\simeq &B_{N}r^{N}:\qquad \qquad \qquad r\rightarrow 0,  \label{As1}
\\
g_{N} &\simeq &1-A_{N}\frac{e^{-\sqrt{2}r}}{\sqrt{r}}:\qquad r\rightarrow
\infty .
\end{eqnarray}
We solved (\ref{Rad}) numerically for $N=1$ and got $B_{1}\approx 0.9$ and $%
\ A_{1}\approx 1.6$.

To find the vortex solution at $N>1$ it is more convenient to use the new
function $v_{N}(r)=g_{N}^{1/N}(r)$ that satisfies the equation: 
\begin{equation}
v_{N}^{\prime \prime }=\frac{v_{N}^{\prime 2}}{v_{N}}-\frac{v_{N}^{\prime }}{%
r}+\frac{1}{N}v_{N}^{2N+1}-\frac{1}{N}v_{N}  \label{RadN}
\end{equation}
and has a linear behavior $\sim B_{N}^{1/N}r$ at $r\rightarrow 0$. The
analytical expression for $g_{N}(r)$ at $N\gg 1$ was obtained in \cite
{Efanov}. In dimensionless units (\ref{dmv}) it is written as: 
\begin{equation}
g_{N}=\left( \frac{r}{r_{N}}\right) ^{n}e^{-\frac{1}{4}(r^{2}-r_{N}^{2})}%
\quad r<r_{N}.  \label{Nvor}
\end{equation}
The size of the vortex core 
\begin{equation}
r_{N}\approx \sqrt{2N}  \label{rad}
\end{equation}
is estimated from that, the almost uniform magnetic field $h_{0}=1$,
distributed inside the vortex area $\pi r_{N}^{2}$ results to the flux $2\pi
N$.

\subsection{\label{slight}Separated vortices and diluted lattice}

The magnetic flux of slightly overlapping vortices can be written as the
superposition of fluxes of separate vortices: 
\begin{equation}
b({\bf r})=\sum_{i}b_{N}({\bf r}-{\bf r}_{i})=\sum_{i}\left(
1-g_{N}^{2}(i)\right) ,  \label{bdil}
\end{equation}
where $g_{N}(i)=g_{N}(\left| {\bf r}-{\bf r}_{i}\right| {\bf )}${\bf \ }is
the solution of (\ref{Rad}). Then, the amplitude of the order parameter is
written as: 
\begin{equation}
\left| \psi ({\bf r,r}_{1}...{\bf r}_{N}{\bf )}\right| ^{2}=1-b({\bf r}%
)=1+\sum_{i}\left( g_{N}^{2}(i)-1\right) .  \label{Psidil}
\end{equation}

\subsection{Vortex bunch}

Consider now the group of $N$ vortices located close to the origin such that 
$\left| {\bf r}_{i}\right| \ll 1$. This vortex bunch can be viewed as the $%
N- $quanta vortex $g_{N}(r)$ with the split core. By direct substitution one
proves that the corresponding solution of equation (\ref{Elstat}) within the
accuracy $O(\max \left| {\bf r}_{i}\right| ^{2})$ is given by:

\begin{eqnarray}
\varphi ({\bf r},{\bf r}_{1}...{\bf r}_{N}) &=&\frac{1}{N}\sum_{i}\varphi
_{N}({\bf r}-{\bf r}_{i})  \label{bunch1} \\
&\approx &\varphi _{N}(r{\bf )+}\frac{1}{2N}\sum_{i}({\bf r}_{i}{\bf \nabla }%
)^{2}\varphi _{N}(r{\bf ),}  \nonumber
\end{eqnarray}
where $\varphi _{N}(r)=\ln \left| g_{N}(r)\right| $, $\varphi ({\bf r},{\bf r%
}_{1}...{\bf r}_{N})=\ln \left| \psi ({\bf r},{\bf r}_{1}...{\bf r}%
_{N})\right| $; the origin being taken in the center of the vortex gravity
such that $\Sigma _{i}{\bf r}_{i}=0$.

With the same accuracy the order parameter is written as:

\begin{equation}
\left| \psi ({\bf r},{\bf r}_{1}...{\bf r}_{N})\right| =g_{N}(r)+\frac{1}{2N}%
g_{N}(r)\sum_{i}({\bf r}_{i}{\bf \nabla })^{2}\ln g_{N}(r).  \label{bunch2}
\end{equation}

\subsection{Dense lattice}

The order parameter of the dense 1-quanta vortex lattice is presented by the
Abrikosov solution close to $H_{c2}$: 
\begin{equation}
\psi _{0}({\bf r)}=A(\overline{b})\theta (\overline{z}\sqrt{\overline{b}\tau
^{\prime \prime }/2\pi },\tau )e^{-\overline{b}y^{2}/2},  \label{dens}
\end{equation}
where: $\ $ $\tau =\tau ^{\prime }+i\tau ^{\prime \prime }=a_{2}e^{i\alpha
}/a_{1}$ is the geometrical parameter of the lattice cell (for square
lattice $\tau =i$; for triangular lattice: $\tau =e^{i\pi /3}$) and $\theta $
is the Jacobi theta-function: 
\begin{equation}
\theta (\overline{z},\tau )=2\sum_{n=0}^{\infty }(-1)^{n}e^{i\pi \tau
(n+1/2)^{2}}\sin [\pi (2n+1)\overline{z}].  \label{Theta}
\end{equation}

Function $\psi _{0}({\bf r)}$ satisfies the linear equation:

\begin{equation}
\overline{b}\nabla ^{2}\ln \left| \psi _{0}\right| =-1+2\pi \sum \delta (%
{\bf r}-{\bf r}_{i}),  \label{Lpsi}
\end{equation}
that close to $H_{c2}$ coincides with BJR equation in a limit $<\left| \psi
\right| ^{2}>\rightarrow 0$, $\overline{b}\rightarrow 1$. To find the
normalization coefficient $A(\overline{b})\sim (1-\overline{b})^{1/2}$ one
should treat the nonlinear part of BJR equation as a perturbation.

The $N$-quanta lattice solution with unit cell area $2\pi N$ can be written
in the analogous way: 
\begin{equation}
\psi _{0}({\bf r)}=A_{N}(\overline{b})\theta ^{N}(\overline{z}\sqrt{%
\overline{b}\tau ^{\prime \prime }/2\pi N},\tau )e^{-\overline{b}y^{2}/2}.
\label{Ndens}
\end{equation}

\subsection{Arbitrary lattice (numerical solution)\label{Sect:Num}}

We performed the numerical integration of BJR equation for square and
triangular vortex lattices with $N=1,2$ in a whole interval of $0<\overline{b%
}<1$ presented it in a more suitable form. First we pick the zeroes of the
order parameter via the special multiplier $\psi _{0}({\bf r)}$ that was
taken as (\ref{dens}) for $N=1$ or as (\ref{Ndens}) for an arbitrary $N$ and
present the order parameter in the form: $\psi ({\bf r)=}\left| \psi _{0}(%
{\bf r)}\right| \cdot \left| \psi ^{\prime }({\bf r)}\right| $. The new
equation for function $\left| \psi ^{\prime }({\bf r)}\right| $ has no
singular $\delta $-function terms and is written as: 
\begin{equation}
\frac{1}{2}\nabla ^{2}\ln \left| \psi ^{\prime }\right| ^{2}=\left| \psi
_{0}\right| ^{2}\left| \psi ^{\prime }\right| ^{2}-(1-\overline{b}).
\label{ne}
\end{equation}
Taking $\varphi ^{\prime }=\ln \left| \psi ^{\prime }\right| $ we present (%
\ref{ne}) in a form:

\begin{equation}
\nabla ^{2}\varphi ^{\prime }=\left| \psi _{0}\right| ^{2}e^{2\varphi
^{\prime }}-(1-\overline{b}).  \label{why}
\end{equation}
This nonlinear Poisson-like equation can have the periodic solution only if
the electroneutrality condition is satisfied: 
\begin{equation}
1-\overline{b}=<\left| \psi _{0}\right| ^{2}e^{2\varphi ^{\prime }}>.
\label{Elnet}
\end{equation}
Taking into account (\ref{Elnet}) and performing the rescaling ${\bf r}%
\rightarrow {\bf r}\cdot (2\pi N/\overline{b})^{1/2}$ we map Eq. (\ref{why})
onto 
\begin{equation}
\nabla ^{2}\varphi ^{\prime }=\Lambda \ (\frac{\left| \psi _{0}({\bf r}%
)\right| ^{2}e^{2\varphi ^{\prime }}}{<\left| \psi _{0}({\bf r})\right|
^{2}e^{2\varphi ^{\prime }}>}-1),  \label{Difeq}
\end{equation}
that is defined for the parallelogram of the fixed area $S=1$ with periodic
boundary conditions. The parameter of the equation 
\begin{equation}
\Lambda =2\pi N\left( \frac{1}{\overline{b}}-1\right)
\end{equation}
varies from $0$ at $H_{c2}$ to $\infty $ at $H_{c1}$.

The problem was solved for the square and triangular vortex lattices with $%
N=1,2$ using the Matlab PDE Toolbox by the finite element method with the
adaptive mesh refinement and with the rapidly converging Gauss-Newton
iterations that were used to account the non-linear right-hand side of (\ref
{Difeq}). For details of the numerical method see Ref. \cite{MatLab}. The
obtained solutions were verified by the direct substitution them back to (%
\ref{Difeq}).

\subsection{Normal-superconducting interface}

The profile of the NS interface is usually considered in two limit cases:
when $\kappa \rightarrow \infty $ or when $\kappa \rightarrow 0$. It
appears, however, that the NS profile can be found {\it exactly} at $\kappa
=1/\sqrt{2}$ by integration of the equation (\ref{Elstat}) that in 1D case
looks like: 
\begin{equation}
\varphi ^{\prime \prime }=e^{2\varphi }-1.  \label{1DBJR}
\end{equation}
The first integral of (\ref{1DBJR}): 
\begin{equation}
(\varphi ^{\prime })^{2}=e^{2\varphi }-2\varphi -1  \label{1intA}
\end{equation}
alternatively can be written as: 
\begin{equation}
(\left| \psi \right| ^{\prime })^{2}=\left| \psi \right| ^{4}-\left| \psi
\right| ^{2}(1+\ln \left| \psi \right| ^{2}).  \label{1intB}
\end{equation}
The integration constant in (\ref{1intA}), (\ref{1intB}) was chosen to
satisfy the NS interface boundary conditions:

\begin{eqnarray}
\left| \psi \right| &=&0,\text{\quad }d\left| \psi \right| /dx=0\text{\quad
when }x\rightarrow -\infty ,  \label{bc} \\
\left| \psi \right| &=&1,\text{\quad }d\left| \psi \right| /dx=0\text{\quad
when }x\rightarrow \infty .  \nonumber
\end{eqnarray}
Further integration of (\ref{1intB}) gives the implicit form of $\left| \psi
(x)\right| $ at the NS interface: 
\begin{equation}
x=\int^{\left| \psi \right| }\frac{dy}{\sqrt{y^{4}-y^{2}(1+\ln y^{2})}}.
\label{NSprof}
\end{equation}

\section{Perturbation theory}

\label{Sect:Perturbation}

To find the most stable configuration of vortices beyond the infinitely
degenerate point $\gamma =0$, $h_{0}=1$ of Bogomolny functional (\ref{Alt})
we construct the secular perturbation functional that acts on the (zero
order) degenerate solutions $\left| \psi ({\bf r},{\bf r}_{1}...{\bf r}%
_{n})\right| $ of (\ref{Elstatalt}) and selects the vortex configuration $%
{\bf r}_{1}...{\bf r}_{n}$ having the lowest energy.

The perturbation for $\gamma $ and $h$ were given already by two last terms
in (\ref{Alt}). To find the perturbation for $t$ one should extend the GL
functional to low temperatures. Tewordt \cite{TewGen,TewPR,TewHc2} and
Newman and Tewordt \cite{TewHc1} were the first who proposed the complete
form of such extension. We will use the analogous functional given in a more
recent publication \cite{Ovch}: 
\begin{eqnarray}
{\cal F} &=&a\left| \Psi \right| ^{2}+\frac{g}{2}\left| \Psi \right|
^{4}+K\left| {\bf D}\Psi \right| ^{2}+\frac{B^{2}}{8\pi }-\frac{BH_{0}}{4\pi 
}  \nonumber \\
&&+\frac{u}{3}\left| \Psi \right| ^{6}+R^{\prime }\left| \Psi \right|
^{2}\left| {\bf D}\Psi \right| ^{2}+R^{\prime \prime }(\nabla \left| \Psi
\right| ^{2})^{2}+P\left| {\bf D}^{2}\Psi \right| ^{2}  \nonumber \\
&&LB^{2}\left| \Psi \right| ^{2}+Qrot^{2}{\bf B.}  \label{EGL}
\end{eqnarray}
The last term $rot^{2}{\bf B}$ was written in \cite{Ovch} in the equivalent
form $-i\,rot{\bf B}(\overline{\Psi }{\bf D}\Psi -\Psi \overline{{\bf D}}%
\overline{\Psi })$.

To account for all the perturbations of the order of $t$ one should assume
that coefficients $u$, $R^{\prime }$, $R^{\prime \prime }$, $P$, $L$, and $Q$
are temperature independent whereas coefficients $\alpha $, $g$ and $K$ are
expanded in $t$ as: 
\begin{eqnarray}
\alpha &=&(\alpha _{1}+\alpha _{2}t)t,  \label{ExtCoef} \\
g &=&g_{0}+g_{1}t,  \nonumber \\
K &=&K_{0}+K_{1}t.  \nonumber
\end{eqnarray}
The microscopic BCS values of these coefficients are given in Appendix.

The combination $g^{1/2}/K$ that enter in (\ref{Alt}) as defined by (\ref
{Param}) parameter $\kappa $ becomes now temperature dependent. We keep a
notation $\kappa =(1+\gamma )/\sqrt{2}$ for the temperature independent part
of (\ref{Param}) and take $\sigma t/\sqrt{2}$ with 
\begin{equation}
\sigma =g_{1}/g_{0}-2K_{1}/K_{0}  \label{sgm}
\end{equation}
as a linear in $t$ contribution to (\ref{Param}). Therefore the third term
in (\ref{Alt}) contains both the perturbation in $\gamma $ and in $t$ and is
written as: 
\begin{equation}
\ (\gamma +\sigma t)(\left| \psi \right| ^{2}-1)^{2}.
\end{equation}

Other perturbation terms of (\ref{EGL}) can be substantially simplified if
one takes into account that they are operating with solutions of BJR
equation. The final form of these terms in dimensionless variables and
corresponding dimensionless coefficients are given in Table I. We present
also the numerical values of these coefficients calculated from microscopic
BCS theory given in Appendix. We comment now, how the perturbation terms
were obtained.

1) The term $P\left| {\bf D}^{2}\Psi \right| ^{2}$ is rewritten in
dimensionless units as: 
\begin{eqnarray}
&&-t\mu \left| 2[(\partial -ia)(\overline{\partial }-i\overline{a})+(%
\overline{\partial }-i\overline{a})(\partial -ia)]\psi \right| ^{2} \\
&=&-t\mu \left| 4(\partial -ia)(\overline{\partial }-i\overline{a})\psi
+2i(\partial \overline{a}-\overline{\partial }a)\psi \right| ^{2}.  \nonumber
\end{eqnarray}
Because of (\ref{Pot}) and (\ref{Second}) the first term in brackets
vanishes and the second term is equal to: 
\begin{equation}
-t\mu b^{2}\left| \psi \right| ^{2}=-t\mu \left| \psi \right| ^{2}(1-\left|
\psi \right| ^{2})^{2}.
\end{equation}

2) The term $R^{\prime }\left| \Psi \right| ^{2}\left| {\bf D}\Psi \right|
^{2}$ is rewritten in dimensionless units as: 
\begin{eqnarray}
&&-2t\rho ^{\prime }\left| \psi \right| ^{2}\left( 2\left| (\partial
-ia)\psi \right| ^{2}+2\left| (\overline{\partial }-i\overline{a})\psi
\right| ^{2}\right) \\
&=&-2t\rho ^{\prime }(4t\left| \psi \right| ^{2}\left| (\overline{\partial }%
-i\overline{a})\psi \right| ^{2}-ia\overline{\partial }\left| \psi \right|
^{4}+i\overline{a}\partial \left| \psi \right| ^{4}).  \nonumber
\end{eqnarray}
The first term in brackets vanishes and two other ones can be integrated by
parts. This leads to: 
\begin{eqnarray}
-t\rho ^{\prime }\left| \psi \right| ^{4}\cdot 2i(\overline{\partial }%
a-\partial \overline{a}) &=&-t\rho ^{\prime }b\left| \psi \right| ^{4} \\
&=&-t\rho ^{\prime }\left| \psi \right| ^{4}(1-\left| \psi \right| ^{2}). 
\nonumber
\end{eqnarray}

3) The term $Qrot^{2}{\bf B}$ in dimensionless units is rewritten as: 
\begin{equation}
-t\tau \,rot^{2}({\bf z}b)=-t\tau (\nabla \left| \psi \right| ^{2})^{2}.
\end{equation}

Multiplying (\ref{Elstatalt}) by $\left| \psi \right| ^{4}$ and integrating
by parts we find that $(\nabla \left| \psi \right| ^{2})^{2}$ can be
substituted as: 
\begin{equation}
(\nabla \left| \psi \right| ^{2})^{2}\rightarrow \left| \psi \right|
^{4}-\left| \psi \right| ^{6},  \label{inden}
\end{equation}
the same substitution was also done for the term $R^{\prime \prime }(\nabla
\left| \Psi \right| ^{2})^{2}$.

Collecting all the above contributions together and omitting the
nonessential constant contribution $1+\gamma +\sigma t-h_{0}$ we came to the
resulting perturbation functional: 
\begin{equation}
f^{\prime }=(h_{0}-h_{c2})\left| \psi \right| ^{2}+(\gamma -c_{4}t)\left|
\psi \right| ^{4}-c_{6}t\left| \psi \right| ^{6},  \label{Secul}
\end{equation}
where parameters$\ c_{i}$ are given in Table II. The instability field: 
\begin{equation}
h_{c2}=1+2\gamma +c_{2}t.  \label{hc2}
\end{equation}
corresponds to the {\it upper critical field} that we discuss below.

Perturbation functional (\ref{Secul}) is the principal result of the present
work. It allows to calculate the properties of superconductor with low $%
\gamma $ and select the most stable vortex configuration at given $\gamma $, 
$h_{0}$ and $t$. Although functional (\ref{Secul}) resembles the extended
form of the GL functional, it is defined for the restricted set of
infinitely degenerate vortex solutions $\left| \psi ({\bf r},{\bf r}_{1}...%
{\bf r}_{n})\right| $ of BJR equation (\ref{Elstatalt}). The important
advantage of the functional (\ref{Secul}) is that, it depends only on few
parameters: $h_{c2}$, $c_{4}$ and $c_{6}$ that are the combinations of the
coefficients of the extended GL functional (\ref{EGL}) as it is given in
Table II. Moreover, it is not necessary to know the coefficients in the
starting functional (\ref{EGL}) at all. These parameters can be considered
as the phenomenological ones. As will be shown in Sec. \ref{Sect:Discussion}
they can be found from experiment.

The occurring vortex state depends on sign and relative strength of the
coefficients $\gamma -c_{4}t$ and $-c_{6}$ that can be both positive and
negative since $t<0$ and parameter $\gamma $ changes the sign when $\kappa $
goes though $1/\sqrt{2}$. The realistic values of $c_{4}$ and $c_{6}$  will
be discussed in Sec. \ref{Sect:Discussion}.

(i) When both $\gamma -c_{4}t$ and $-c_{6}t$ are positive the magnetic
behavior of the superconductor corresponds to the generic scenario for a
superconductor of type II. The dense vortex lattice (\ref{dens}) appears
continuously from the normal state $\left| \psi \right| =0$ at upper
critical field $h_{c2}$ when the quadratic term $(h_{0}-h_{c2})\left| \psi
\right| ^{2}$ in (\ref{Secul}) becomes unstable. The amplitude $\Delta
^{2}=<\left| \psi ({\bf r})\right| ^{2}>$ of the vortex state and the
intervortex distance increase with decreasing of the applied field. Below
the {\it low critical field }$h_{c1}$ that will be calculated in Sec.\ref
{Sect:lcf} all the vortices continuously leave the superconductor and the
uniform Meissner state with $\left| \psi \right| =1$ becomes stable.

(ii) When both $\gamma -c_{4}t$ and $-c_{6}t$ are negative, the functional (%
\ref{Secul}) corresponds to a type-I superconductor. There are only two
competing local minima of (\ref{Secul}): the normal state with $\left| \psi
\right| =0$ and energy 
\begin{equation}
f_{n}^{\prime }=0,  \label{NormalHc1}
\end{equation}
and the uniform superconducting Meissner state with $\left| \psi \right| =1$
and energy: 
\begin{equation}
f_{s}=h_{0}-h_{c2}+\gamma -(c_{4}+c_{6})t.  \label{SCHc1}
\end{equation}
The discontinuous transition between them occurs at the{\it \ thermodynamic
critical field }$h_{c}$ 
\begin{equation}
h_{c}=1+\gamma +(c_{2}+c_{4}+c_{6})t.  \label{hc}
\end{equation}
that is found by equating (\ref{NormalHc1}) and (\ref{SCHc1}).

(iii) We investigate the case when coefficients $\gamma -c_{4}t$ and $%
-c_{6}t $ have different signs in Sec. \ref{Sect:Discussion}. It will appear
that, depending on the situation, both N-V and V-S transition can be either
continuous (as in conventional superconductors with $\kappa \gg 1/\sqrt{2}$)
or discontinuous. This situation is accessible experimentally either by
variation of $\gamma $ or by variation of $t$.

\section{Vortex state: energy and critical fields}

\label{Sect:VortEnergy}

\subsection{\label{evl}Energy of the vortex lattices and higher critical
field}

The energy of the regular $N$-quanta lattice (\ref{Secul}) can be written in
terms of the amplitude of the order parameter $\Delta ^{2}=<\left| \psi
\right| ^{2}>$ as: 
\begin{eqnarray}
f^{\prime } &=&(h_{0}-h_{c2})\Delta ^{2}  \nonumber \\
&&+(\gamma -c_{4}t)\beta _{4}(\Delta )\Delta ^{4}-c_{6}t\beta _{6}(\Delta
)\Delta ^{6},  \label{princ}
\end{eqnarray}
where $\beta _{n}^{(N)}(\Delta )$ are the structural factors: 
\begin{equation}
\beta _{n}^{(N)}(\Delta )=\frac{<\left| \psi \right| ^{n}>}{<\left| \psi
\right| ^{2}>^{n/2}},  \label{betaall}
\end{equation}
that depend both on the amplitude $\Delta $ and on the lattice geometry.

Minimization of (\ref{princ}) over $\Delta $ gives the complete information
about thermodynamic and magnetic properties of vortex lattice in a
superconductor with $\kappa \sim 1/\sqrt{2}$, provided the dependencies $%
\beta _{n}^{(N)}(\Delta )$ are known. We found $\beta _{n}^{(N)}(\Delta )$
in the whole region of $\Delta \ $($0<\Delta <1$) using the numerical
solutions of (\ref{Elstatalt}) for square and triangular vortex lattices
with $N=1,2$ outlined in Sec. \ref{Sect:Num}. The results are shown in Fig.1
as function of magnetic induction 
\begin{equation}
\overline{b}=1-\Delta ^{2}.  \label{ami}
\end{equation}
The values of $\beta _{n}^{(1,2)}$ at $\Delta \rightarrow 0$ (i.e. in
vicinity of $h_{c2}$) are given in Table III. Parameter $\beta
_{4}^{(N)}(0)\ $corresponds to the parameter $\beta $ introduced in the
original publication of Abrikosov \cite{Abr}. Close to $h_{c1}$ (where $%
\Delta \rightarrow 1$) the factors $\beta _{n}^{(N)}$ are tending to $1$.
The \ corresponding asymptotic expression will be given in Sec. \ref
{Sect:lcf}.

Functional (\ref{princ}) close to $h_{c2}$\ can be interpreted as a Landau
expansion of the vortex state energy over the amplitude $\Delta $. When $%
\gamma >c_{4}t$ the quartic in $\Delta $ term is positive and the
conventional second order transition occurs at $h_{c2}$. When $\gamma <c_{4}t
$ the transition occurs in a discontinuous way either to the
finite-amplitude vortex state or directly to the Meissner state. The
concrete realization of this transition depends on the relative values of $%
c_{6}$ and $c_{4}$ and will be discussed in Sec. \ref{Sect:Discussion}.

Condition: 
\begin{equation}
\gamma -c_{4}t=0  \label{fourthneg}
\end{equation}
defines the tricritical point $T_{2}$ where the discontinuity of N-V
transition appears. The discontinuity field $h_{c2}^{\ast }$ is large than
the critical field $h_{c2}$.

Functional (\ref{princ}) can be alternatively written in terms of $\overline{%
b}$ as: 
\begin{eqnarray}
f^{\prime } &=&(2\gamma +c_{2}t)(h_{0}-1)+(h_{0}-h_{c2})(h_{c2}-\overline{b})
\label{mgn} \\
&&+(\gamma -c_{4}t)\beta _{4}(\overline{b})(h_{c2}-\overline{b}%
)^{2}-c_{6}t\beta _{6}(\overline{b})(h_{c2}-\overline{b})^{3}.  \nonumber
\end{eqnarray}
Minimization of $f^{\prime }$ over $\overline{b}$ gives the induction $%
\overline{b}(h_{0})$ and the lattice energy $f^{\prime }(h_{0})$. Comparing
the energies of square and triangular lattices with $N=1,2$ at given $h_{0}$
($h_{c1}<h_{0}<h_{c2}$) we established that {\it one-quanta triangular
vortex lattice} always possess the lowest energy. However close to $h_{c1}$
the energies of different lattices coincide within the calculation accuracy
and this conclusion becomes less certain.

\subsection{Diluted lattice, low critical field and vortex interaction}

\label{Sect:lcf}

\subsubsection{Energy of the diluted vortex lattice}

We consider now the diluted lattice of slightly overlapping $N$-quanta
vortices assuming that the distance $l$ between them is much larger then the
coherence length (i.e. $l\gg 1$). Such limit usually occurs close to the low
critical field $h_{c1}$. In this approximation the energy of the system is
written as: 
\begin{equation}
f^{\prime }=f_{s}^{\prime }+\frac{\overline{b}}{2\pi N}%
\varepsilon _{N}+\frac{m}{2}\frac{\overline{b}}{2\pi N}U_{int}(l)-h_{0}%
\overline{b},  \label{dillatenfirst}
\end{equation}
where the background energy of the uniform Meissner state $f_{s}^{\prime }$
is given by (\ref{SCHc1}), $\ \varepsilon _{N}$ \ is the one-vortex energy, $%
\overline{b}/2\pi N$ is the density of vortices and $-h_{0}\overline{b}$ is
the interaction of the vortex with an external field. The term $U_{int}(l)$
represents the interaction between the nearest neighbor vortices. The factor 
$\ m$ gives the lattice coordination number: $m=6$ for the triangular
lattice and $m=4$ for the square lattice. The inter-vortex distance $l$ is
uniquely related with the vortex concentration $2\pi N/\overline{b}$ and
geometry of the lattice as: 
\begin{equation}
l_{\square }=(2\pi N/\overline{b})^{1/2},\quad l_{\Delta }=(4\pi N/\overline{%
b}\sqrt{3})^{1/2}.  \label{lotb}
\end{equation}
The type of the V-S transition depends on sign of the long-range vortex
interaction $U_{int}(l)$ that, as will be shown below, can be both repulsive
and attractive.

When $U_{int}(l)>0$ the situation is the same as for a superconductor of
type-II: the V-S transition occurs in a continuous way at the low critical
field $h_{c1}$ that is calculated from the vortex energy $\varepsilon _{N}$.
The latter can be written on the basis of (\ref{Secul}) as: 
\begin{equation}
\varepsilon _{N}=2\pi N[(1+2\gamma +c_{2}t)-\zeta _{4}^{(N)}(\gamma
-c_{4}t)+\zeta _{6}^{(N)}c_{6}t].  \label{EV}
\end{equation}
The structural factors for the $N$-quanta vortex: 
\begin{equation}
\zeta _{n}^{(N)}=\frac{1}{N}\int_{0}^{\infty }(1-g_{N}^{n})rdr  \label{vstrf}
\end{equation}
are found from integration of the numerical solution $g_{N}(r)$ of Eq. (\ref
{Rad}) and are given in Table III. Note also that: 
\begin{equation}
\zeta _{2}^{(N)}=\frac{\overline{b}}{2\pi N}=1\text{.}
\end{equation}
Factors $\zeta _{n}^{(N)}$ for large $N$ will be calculated in Sec. \ref
{NSen}. Lattice factors $\beta _{n}^{(N)}$ (\ref{betaall}) can be expressed
via $\zeta _{n}^{(N)}$ as: 
\begin{equation}
\beta _{n}^{(N)}=\frac{1-\zeta _{n}^{(N)}\overline{b}}{(1-\overline{b})^{n/2}%
}\quad (\overline{b}\rightarrow 0).
\end{equation}
The $N$-quanta vortices penetrate into a sample when the positive energy $%
\varepsilon _{N}$, required for the vortex \ creation is compensated by the
negative magnetic contribution $-h_{0}\overline{b}$ i.e. above the critical
field: 
\begin{eqnarray}
h_{c1}^{(N)} &=&\frac{\varepsilon _{N}}{2\pi N}=1+(2-\zeta _{4}^{(N)})\gamma
\label{hc1} \\
&&+(c_{2}+\zeta _{4}^{(N)}c_{4}+\zeta _{6}^{(N)}c_{6})t.  \nonumber
\end{eqnarray}
The low critical field is defined as the lowest field when penetration of
vortices becomes favorable: 
\begin{equation}
h_{c1}=\min \{h_{c1}^{(N)}\}_{N}.
\end{equation}

It appears that only the 1-quanta vortices can appear in a continuous way
since condition of formation of 2-quanta vortices written as $%
h_{c1}^{(2)}<h_{c1}^{(1)}$ or as: 
\begin{equation}
\gamma >(c_{4}+1.89c_{6})t  \label{2l1}
\end{equation}
is weaker then the condition of continuity of V-S transition: $U_{int}(l)>0$%
, derived below (inequality (\ref{repel})).

Vortices penetrate inside a superconductor until the repulsive interaction
counterbalances the energy win. The penetrated flux is determined by
minimization of (\ref{dillatenfirst}) over $\overline{b}$ that alternatively
can be written as: 
\begin{equation}
 f^{\prime }=f_{s}^{\prime }+[h_{c1}-h_{0}+\frac{m}{4\pi N}%
U_{int}(l(\overline{b}))]\overline{b},  \label{ffl}
\end{equation}
where dependence $l(\overline{b})$ is given by (\ref{lotb}).

When $U_{int}(l)<0$, the transition from the Meissner phase occurs either to
the finite-density vortex state or directly to the normal-metal state in a
discontinuous way that is manifested by the jump of magnetization. The
detailed scenario of the transition depends on the energy balance between
these three phases and will be discussed in Sec. \ref{Sect:Discussion}.

The situation is simplified however near the tricritical point $T_{1}$ where
the long-range part of $U_{int}(l)$ changes its sign from positive to
negative.\ As will be shown in Sec. \ref{LR} the short-range vortex
interaction in this region is still repulsive. The minimum of $U_{int}(l)$
lies at $l\gg 1$ and one can apply the nearest-neighbor approximation (\ref
{ffl}). The discontinuity field $h_{c1}^{\ast }$ is smaller than $h_{c1}$.

\subsubsection{\label{LR}Vortex interaction}

The vortex interaction is known to be repulsive in a type-II superconductor (%
$\gamma \gg 0$) and attractive in a type-I superconductor ($\gamma \ll 0$).
In this Section we calculate the interaction energy $U_{int}(l)$ of two $N$%
-quanta vortices located at ${\bf r}_{1,2}=\pm {\bf l}/2$ at the
intermediate values of $\gamma $. The numerical part of this problem is
based on solution of the BJR equation (\ref{Elstatalt}) with the right-hand
term $2\pi N\delta ({\bf r}-{\bf r}_{1})+2\pi N\delta ({\bf r}-{\bf r}_{2})$
and on substitution of this solution into the perturbation functional (\ref
{Secul}). We give the analytical treatment of this problem in cases of
slightly overlapped ($l\gg 1$) and of strongly overlapped ($l\ll 1$)
vortices. As the result we obtain that vortices begin to attract each other
at large distances when:

\begin{equation}
\gamma <(c_{4}+3c_{6})t.  \label{repel}
\end{equation}
Below this instability the vortex interaction has the long-range attractive
and short-range repulsive character and vortices form a bounded state.
Inequality (\ref{repel}) presents also the condition of discontinuity of VS
transition.

With decrease of $\gamma $ the equilibrium distance $l_{0}$ varies from
infinity to zero. Below another instability point at

\begin{equation}
\gamma <(c_{4}+1.5\frac{\zeta _{8}^{(N)}-\zeta _{6}^{(N)}}{\zeta
_{6}^{(N)}-\zeta _{4}^{(N)}}c_{6})t  \label{attr}
\end{equation}
the interaction is purely attractive and vortices are stuck together with
formation of $2N$-quanta vortex. The results about the short-range vortex
interaction can not be directly applied to study the vortex lattice since
the nearest-neighbor approximation (\ref{ffl}) is not applicable at low
vortex separation $l$ .

The given below calculations of the long-range vortex interaction are
compatible with calculations of the vortex interaction given in \cite{Jac123}
in a different way. Consider for simplicity the case of two 1-quanta
vortices. When the distance between vortices is large ($l\gg 1$) it is more
suitable to describe the vortices in terms of slightly overlapping magnetic
fluxes produced by these vortices: 
\begin{equation}
b_{\pm }=1-g_{1}^{2}(\left| {\bf r\pm l}/2\right| ),  \label{twofar}
\end{equation}
as was discussed in Sec. \ref{slight}.

The vortex energy is provided by the terms $\left| \psi \right| ^{2}$, $%
\left| \psi \right| ^{4}$ and $\left| \psi \right| ^{6}$ in the functional (%
\ref{Secul}) that can be evaluated as: 
\begin{equation}
\left| \psi \right| ^{2}=1-b_{+}-b_{-},  \label{p1}
\end{equation}
\begin{gather}
\left| \psi \right| ^{4}=(1-b_{+}-b_{-})^{2}  \label{p4} \\
=(1-b_{+})^{2}+(1-b_{+})^{2}-1+2b_{+}b_{-},  \nonumber
\end{gather}
and 
\begin{gather}
\left| \psi \right| ^{6}=(1-b_{+}-b_{-})^{3}  \label{p6} \\
=(1-b_{+})^{3}+(1-b_{+})^{3}-1+6b_{+}b_{-}-3b_{-}^{2}b_{+}-3b_{+}^{2}b_{-}. 
\nonumber
\end{gather}

Only $\left| \psi \right| ^{4}$ and $\left| \psi \right| ^{6}$ terms contain
the interaction parts: $b_{+}b_{-}$, $b_{+}^{2}b_{-}$ and $b_{+}b_{-}^{2}$.
\ With the help of (\ref{As1}) the overlapping contribution \ $<b_{+}b_{-}>$
can be estimated with exponential accuracy as $u(l)e^{-4l}$ where $u(l)$ is
a slow function of $l$. This term is more important at $l\gg 1$ then
decaying as $e^{-6l}$ terms $<b_{+}b_{-}^{2}>,<b_{+}^{2}b_{-}>$.

Substituting the contributions $\ <b_{+}b_{-}>$ from (\ref{p4}), (\ref{p6})
to (\ref{Secul}) one gets the long-range interaction energy per one vortex:

\begin{equation}
U_{int}(l)=[\gamma -(c_{4}+3c_{6})t]\cdot u(l)e^{-4l}.  \label{Inten}
\end{equation}
This interaction is attractive when condition (\ref{repel}) is satisfied.

Consider now the short-range part of the vortex interaction. When vortices
are located close to each other ($l\ll 1$), their order parameter is given
by (\ref{bunch2}):

\begin{equation}
\left| \psi ({\bf r})\right| =g_{2}(r)+\frac{1}{8}g_{2}(r)({\bf l\nabla }%
)^{2}\ln g_{2}(r).  \label{twoclose}
\end{equation}
To calculate the vortex energy one should estimate:

\begin{gather}
\int \left| \psi ({\bf r})\right| ^{n}d^{2}{\bf r}\approx 2\pi \int
g_{2}^{n}(r)rdr \\
+\frac{n}{8}\int g_{2}^{n}(r)({\bf l\nabla })^{2}\ln g_{2}(r)d^{2}{\bf r}. 
\nonumber
\end{gather}
Since $g_{2}(r)$ is an axisymmetric function, the operator $({\bf l\nabla }%
)^{2}$ can be substituted by $\frac{l^{2}}{2}{\bf \nabla }^{2}$. Taking
finally into account the BJR equation ${\bf \nabla }^{2}\ln
g_{2}=g_{2}^{2}-1 $ one gets:

\begin{eqnarray}
\int \left| \psi ({\bf r})\right| ^{n}d^{2}{\bf r} &\approx &2\pi \int
g_{2}^{n}(r)rdr  \label{clso} \\
&&+\frac{\pi nl^{2}}{4}(\zeta _{n+2}^{(2)}-\zeta _{n}^{(2)}).  \nonumber
\end{eqnarray}

The interaction energy is calculated on the base of (\ref{Secul}) with
respect to the state where vortex cores coincide. With help of (\ref{clso})
one gets the short-range interaction energy:

\begin{equation}
U_{int}(l)=[(\gamma -c_{4}t)(\zeta _{6}^{(2)}-\zeta
_{4}^{(2)})-1.5c_{6}t(\zeta _{8}^{(2)}-\zeta _{6}^{(2)})]\cdot \pi l^{2},
\end{equation}
This interaction is attractive when condition (\ref{attr}) is satisfied.

\subsection{\label{NSen}Energy of the normal-superconducting interface}

The profile of the NS interface is given by (\ref{NSprof}). Basing on our
perturbation approach we calculate the NS interface energy at $h_{0}=h_{c}$
as: 
\begin{equation}
\sigma _{ns}=-(\gamma -c_{4}t)\alpha _{4}+c_{6}t\alpha _{6},
\end{equation}
where the structural factors $\alpha _{n}$ are defined as: 
\begin{equation}
\alpha _{n}=\int_{-\infty }^{\infty }(\left| \psi \right| ^{2}-\left| \psi
\right| ^{n})dx.  \label{alphan}
\end{equation}
With help of (\ref{1intA}) these factors can be present in the form of
definite integrals: 
\begin{gather}
\alpha _{n}=\int_{-\infty }^{\infty }(e^{2\varphi }-e^{n\varphi
})dx=-\int_{\varphi =-\infty }^{\varphi =0}\frac{e^{2\varphi }-e^{n\varphi }%
}{1-e^{2\varphi }}d\varphi _{x}^{\prime } \\
=\int_{0}^{\infty }\frac{e^{-2\eta }-e^{-n\eta }}{\sqrt{e^{-2\eta }+2\eta -1}%
}d\eta ,  \nonumber
\end{gather}
and calculated numerically: 
\begin{equation}
\alpha _{4}\simeq 0.55,\quad \alpha _{6}\simeq 0.85,\quad \alpha _{8}\simeq
1.06.  \label{alphas}
\end{equation}

One can now express the vortex structural factors $\zeta _{n}^{(N)}$ (\ref
{vstrf}) for $N\gg 1$ via the NS interface factors $\alpha _{n}$. The $N$%
-quanta vortex \ (\ref{Nvor}) can be viewed as the cylindric normal-state
domain of radius $r_{N}\approx \sqrt{2N}$\ surrounded by the NS interface.
The domain energy is written as: 
\begin{equation}
\varepsilon _{N}=\pi r_{N}^{2}h_{c}+2\pi r_{N}\sigma _{ns},  \label{estm}
\end{equation}
where $\pi r_{N}^{2}h_{c}$ is the energy of the condensate break inside the
domain and $2\pi r_{N}\sigma _{ns}$ is the domain wall energy. Comparison of
(\ref{estm}) and (\ref{EV}) gives: 
\begin{equation}
\zeta _{n}^{(N)}=1+\frac{\alpha _{n}}{\sqrt{N}}\quad (N\gg 1).
\label{largeN}
\end{equation}
Interesting to note that formula (\ref{largeN}) can be extrapolated to small 
$N$ with an accuracy $5-8\%$.

\section{H-T phase diagram}

\label{Sect:Discussion}We are now in a stage to discuss the properties of $%
H-T$ diagram of a superconductor with $\kappa \sim 1/\sqrt{2}$. (We use
again the dimensional variables.) Partially this question was considered in 
\cite{Jac123} on the base of Neumann-Tewordt extension of GL equations to
the low temperature. The advantage of our approach is that, it allows to get
the structure of $H-T$ diagram in a unified way from a simple perturbation
functional (\ref{Secul}). This functional depends on three driving
parameters: $h_{0}$, $t$, and $\gamma $ that are controlled by experimental
conditions and on three phenomenological parameters: $h_{c2}$, $c_{4}$ and $%
c_{6}$ that can be found from experimentm, basing on the relations:

\begin{equation}
\frac{H_{c2}}{H_{c}}=1+\gamma -(c_{4}+c_{6})t,  \label{e1}
\end{equation}
\begin{equation}
\left( \frac{dM}{dH}\right) _{H=H_{c2}}=\frac{1}{8\pi \beta _{4}(\gamma
-c_{4}t)},
\end{equation}
\begin{equation}
\frac{H_{c}}{H_{c1}}=1+(\zeta _{4}-1)\gamma +[(\zeta _{4}-1)c_{4}+(\zeta
_{6}-1)c_{6}]t,
\end{equation}
extracted from (\ref{hc2}), (\ref{hc}), (\ref{mgn}) and (\ref{hc1}). The
first value was also called as $\kappa _{1}(T)$, the second one as $4\pi
\beta _{4}(2\kappa _{2}^{2}(T)-1)$ $\ $and the third one as $2\kappa
_{3}/\ln \kappa _{3}$ \cite{Saint}.

We extracted the parameters $c_{4}$ and $c_{6}$ from magnetic measurements
in TaN \cite{(Ta-Nb)Nmagn} and got: $c_{4}\simeq 0.30$ and $c_{6}\simeq -0.15
$. These parameters can be also estimated theoretically from the microscopic
BCS-expression for coefficients of the extended GL functional \cite{Ovch}.
Calculations presented in Appendix and in Tables I and II give $c_{4}=0.5$, $%
c_{6}=-0.09$ for clean superconductor and $c_{4}=0.1$, $c_{6}=0.01$ for the
dirty superconductor. Although these estimations do not take into account
the anisotropy of TaN and the electron-phonon retardation effects in BCS
theory, they give the correct idea about the magnitude of coefficients $c_{4}
$ and $c_{6}$. We assume further that $c_{4}$ varies from $0.1$ to $0.5$ and 
$c_{6}$ from $-0.2$ to $0.01$; the negative value of $c_{6}$ being more
probable.

The $H-T$ diagram of low-$\gamma $ superconductor at given $c_{4}$ and $%
c_{6} $ can be obtained from comparison of relative location of critical
fields (\ref{hc2}), (\ref{hc}), (\ref{hc1}) and characteristic critical
points that were calculated in Sec.\ref{Sect:VortEnergy} and that are
resumed in Table IV. To avoid the narrowness of the mixed state region and
to clearly demonstrate the details of the $H-T$ diagram we trace it in the
specially normalized coordinates $T/T_{c}$ and $H/H_{c}$ where $H_{c}$
linearly depends on $T$:

\begin{equation}
H_{c}=H_{c}^{\prime }(T_{c}-T),
\end{equation}
with $H_{c}^{\prime }=(4\pi /g)^{1/2}\alpha _{1}/T_{c}$ (\ref{Hc}). Topology
of the $H-T$ diagram depends on the relative strength of coefficients $c_{6}$
and $c_{4}$. Three possible scenario of N-S transition can be distinguished.

i) Fig.2 corresponds to $c_{6}$,$c_{4}>0$ and to $\gamma <0$. Between $T_{c}$
and triple point $L_{1}$ defined by condition $H_{c}(T)=H_{c1}(T)$ or: 
\begin{equation}
\gamma =(c_{4}+1.75c_{6})t,
\end{equation}
the superconductor behaves like superconductor of type-I i.e. the
discontinuous N-S transition occurs at critical field $H_{c}$. On the left
of $L_{1}$ the N-S transition occurs via an intermediate vortex state. The
V-S transition occurs in a continuous way at $H=H_{c1}$ like in type-II
superconductor. The N-V transition has a discontinuous character close to $%
L_{1}$. The discontinuity line $H_{c2}^{\ast }$ terminates in the defined by
(\ref{fourthneg}) tricritical point $T_{2}$ where the forth-order term in
functional (\ref{princ}) becomes positive. On the left of $T_{2}$ the N-V
transition occurs in a continuous way at $H=H_{c2}$ $\ $and\ the
superconductor behaves like conventional type-II superconductor. When $%
\gamma $ decreases, points $L_{1}$ and $T_{2}$ are shifted to low
temperatures and the superconductor becomes superconductor of type I in a
whole temperature region. \ When $\gamma $ increases, points $L_{1}$ and $%
T_{2}$ are shifted to $T_{c}$. At positive $\gamma $ the superconductor has
a type II behavior.

ii) Fig.3 corresponds to $c_{4}>0$, $0>c_{6}>-c_{4}/3$ and to $\gamma <0$.
Similar to the case i), the vortex state appears on the left of the triple
point $L_{2}$ that is defned by condition $H_{c2}(T)=H_{c}(T)$, or: 
\begin{equation}
\gamma =(c_{4}+c_{6})t.
\end{equation}
The N-V transition has a continous character and occures at $H=H_{c2}$. The
type of the V-S tranition is provided by the location of tricritical point $%
T_{1}$ where the long-range vortex interaction chages the sign (condition (%
\ref{repel})). The V-S transition is discontinuous between $L_{2}$ and $%
T_{1} $ at $H=H_{c1}^{\ast }$ and continuous on the left of $T_{1}$ at $%
H=H_{c1}$.When $\gamma $ decreases the superconductor transforms to
superconductor of type-I whereas when $\gamma $ increases above zero it
becomes a superconductor of type-II.

iii) Fig. 4 corresponds to $c_{4}>0$, $-c_{4}/3>c_{6}$. This case
corresponds to experimental situation in TaN \cite{(Ta-Nb)Nmagn}.When $%
\gamma $ is slightly less then zero (Fig.4a), the $H-T$ diagram is obtained
from the diagram of case ii) by shift of the tricritical point $T_{1}$ to
the region of negative temperatures. Like in previous cases, the
superconductor transforms to a superconductor of type-I with decreasing of $%
\gamma $. When $\gamma $ increases, point $L_{2}$ goes to $T_{c}$ and
disappears at $\gamma =0$. When $\gamma $ becomes positive (Fig.4b) the V-S
transition is continuous between $T_{c}$ and tricritical point $T_{1}$ that
appears at $T_{c}$ at $\gamma =0$ and is moving to the region of low
temperatures with increasing of $\gamma $. On the left of $T_{1}$ the V-S
transition has a discontinous character. At large $\gamma $ the phase
diagram transforms to conventional $H-T$ diagram of superconductor of
type-II.

\acknowledgments

I am grateful to Y.N. Ovchinnikov, V. Geshkenbein and J. Blatter for helpful
discussions of the theoretical aspects and to L.Y.Vinnikov for explication
of the experimental situation. Especially I thank H. Capellmann for his
hospitality during my stay in RWTH-Aachen. The work was done in part at the
Federal University of Minas Gerais, Brazil and at the Eidgen\"{o}ssische
Technische Hochschule, Z\"{u}rich.

\appendix

\section*{Microscopic parameters}

The microscopic BCS parameters of the extended GL functional (\ref{EGL})
were calculated in \cite{Ovch}. The first three coefficients do not depend
on the purity of superconductor: 
\begin{eqnarray}
\alpha &=&\nu \ln \frac{T}{T_{c}},\qquad  \label{BCS} \\
\frac{g}{2} &=&\nu \frac{7\zeta (3)}{16}\frac{1}{(\pi T)^{2}},\qquad \frac{u%
}{3}=-\frac{\nu }{2}\frac{31\zeta (5)}{64}\frac{1}{(\pi T)^{4}}.  \nonumber
\end{eqnarray}
(Here $\nu =mp_{F}/2\pi ^{2}\hbar ^{3}$ is the density of states).

Other coefficients depend on the quality of material and can be calculated
in two limit cases:

\subsubsection{Clean limit}

\begin{eqnarray}
K &=&\nu \frac{7\zeta (3)}{48}\frac{v^{2}\hbar ^{2}}{(\pi T)^{2}},\qquad P=-%
\frac{\nu }{20}\frac{31\zeta (5)}{64}\frac{\hbar ^{4}v^{4}}{(\pi T)^{4}},
\label{Clean} \\
R^{\prime } &=&-\frac{\nu }{3}\frac{31\zeta (5)}{64}\frac{v^{2}\hbar ^{2}}{%
(\pi T)^{4}},\qquad R^{\prime \prime }=-\frac{\nu }{12}\frac{31\zeta (5)}{64}%
\frac{v^{2}\hbar ^{2}}{(\pi T)^{4}},  \nonumber \\
Q &=&-\frac{1}{35\pi }\frac{31\zeta (5)}{64\zeta (3)}\frac{\hbar ^{2}v^{2}}{%
(\pi T)^{2}},\qquad L=-\frac{\nu }{5}\frac{31\zeta (5)}{64}\frac{e^{2}}{%
c^{2}\hbar ^{2}}\frac{\hbar ^{4}v^{4}}{(\pi T)^{4}}.  \nonumber
\end{eqnarray}

\subsubsection{Dirty limit}

\begin{eqnarray}
K &=&\nu \frac{\pi ^{2}}{48}\frac{v^{2}\hbar ^{2}}{s_{1}\pi T},\qquad P=-\nu 
\frac{7\zeta (3)}{12\cdot 48}\frac{\hbar ^{4}v^{4}}{s_{1}^{2}(\pi T)^{2}},
\label{Dirty} \\
R^{\prime } &=&-\nu \frac{\pi ^{4}}{12\cdot 48}\frac{v^{2}\hbar ^{2}}{%
s_{1}(\pi T)^{3}},\qquad R^{\prime \prime }=-\nu \frac{\pi ^{4}}{48^{2}}%
\frac{v^{2}\hbar ^{2}}{s_{1}(\pi T)^{3}},  \nonumber \\
Q &=&-\frac{1}{80\pi }\frac{\hbar ^{2}v^{2}}{s_{1}s_{2}},\qquad L=-\nu \frac{%
\pi ^{2}}{80}\frac{e^{2}}{c^{2}\hbar ^{2}}\frac{\hbar ^{4}v^{4}}{%
s_{1}^{2}s_{2}\pi T},  \nonumber
\end{eqnarray}
where parameters $s_{1}$ and $s_{2}$ are the functions of the scattering
times $\tau $, $\tau _{1}$, $\tau _{2}$ in the $s,p$ and $d$ channels: 
\begin{equation}
s_{1}=\frac{\hbar }{2\tau _{tr}}=\frac{\hbar }{2}(\frac{1}{\tau }-\frac{1}{%
\tau _{1}}),\qquad s_{2}=\frac{\hbar }{2}(\frac{1}{\tau }-\frac{1}{\tau _{2}}%
).  \label{s12}
\end{equation}

From (\ref{BCS}),(\ref{Clean}) and (\ref{Dirty}) we calculate the
microscopic expressions for the defined by (\ref{sgm}) parameter $\sigma $: 
\begin{equation}
\sigma _{cl}=1,\ \ \ \ \ \ \sigma _{d}=0,
\end{equation}
and for the GL parameter $\kappa $: 
\begin{equation}
\kappa _{cl}=\frac{3}{(7\pi \zeta (3)\nu )^{1/2}}\frac{\pi Tc}{\left|
e\right| \hbar v^{2}},\ \ \ \ \ \ \kappa _{d}=\frac{3(7\zeta (3))^{1/2}}{\pi
^{2}(\pi \nu )^{1/2}}\frac{cs_{1}}{\left| e\right| \hbar v^{2}}
\end{equation}
in clean and in dirty superconductors.

%\[ B=-H_{c}^{\prime }[b-\gamma +(0.5-\sigma )t]t \]

\em

\newpage 

\widetext

\begin{table}[tbp]
\caption{
The terms of the extended GL functional $F$, their dimensionless counterparts
and coefficients in the perturbation functional $f$. Two last 
columns give the microscopic BCS values of coefficients   
for clean and dirty superconductors.}
\begin{tabular}{lllll}
$F$ & $f$ & coefficient & clean limit & dirty limit \\ \hline
$\frac{u}{3}\left| \Psi \right| ^{6}$ & $-t\upsilon \left| \psi \right| ^{6}$
& $\upsilon =2\alpha _{1}\kappa ^{2}u/3g^{2}$ & \multicolumn{2}{c}{$\upsilon
=-\vartheta /2\approx -0.23$ \ \ \ \ \ \ \ \ \ } \\ 
$R^{\prime }\left| \Psi \right| ^{2}\left| {\bf D}\Psi \right| ^{2}$ \ \ \ \
\ \  & $-t\rho ^{\prime }\left| \psi \right| ^{4}(1-\left| \psi \right|
^{2}) $ \ \ \ \ \ \  & $\rho ^{\prime }=\alpha _{1}R^{\prime }/2gK$ \ \ \ \
\ \ \ \ \  & $\rho ^{\prime }=-\vartheta \approx -0.45$ \ \ \ \ \ \ \ \ \  & 
$\rho ^{\prime }=-\pi ^{2}/21\zeta (3)\approx -0.39$ \\ 
$R^{\prime \prime }(\nabla \left| \Psi \right| ^{2})^{2}$ & $-t\rho ^{\prime
\prime }\left| \psi \right| ^{4}(1-\left| \psi \right| ^{2})$ & $\rho
^{\prime \prime }=\alpha _{1}R^{\prime \prime }/gK$ & $\rho ^{\prime \prime
}=-\vartheta /2\approx -0.23$ & $\rho ^{\prime \prime }=-\pi ^{2}/42\zeta
(3)\approx -0.20$ \\ 
$P\left| {\bf D}^{2}\Psi \right| ^{2}$ & $-t\mu \left| \psi \right|
^{2}(1-\left| \psi \right| ^{2})^{2}$ & $\mu =\alpha _{1}P/2\kappa ^{2}K^{2}$
& $\mu =-9\vartheta /5\approx -0.82$ & $\mu =-28\zeta (3)/\pi ^{4}\approx
-0.35$ \\ 
$LB^{2}\left| \Psi \right| ^{2}$ & $-t\lambda (\left| \psi \right|
^{2}-\left| \psi \right| ^{4})$ & $\lambda =4\pi \alpha _{1}L/g$ & $\lambda
=-9\vartheta /5\approx -0.82$ & $\lambda =-36T/5\pi s_{2}\rightarrow 0$ \\ 
$Qrot^{2}{\bf B}\qquad $ & $-t\tau \left| \psi \right| ^{4}(1-\left| \psi
\right| ^{2})$ & $\tau =4\pi \alpha _{1}Q/2\kappa ^{2}K$ & $\tau
=-3\vartheta /5\approx -0.27$ & $\tau =-12T/5\pi s_{2}\rightarrow 0$ \\
\end{tabular}
\tablenotetext{ $\vartheta =31\zeta (5)/49\zeta ^{2}(3)\approx 0.454$}
\end{table}

\bm

%%%%%%%%%%%%%%%%%%%%%%%%%%%%%%%%%%%%%%%%%

\narrowtext

%%%%%%%%%%%%%%%%%%%%%%%%%%%%%%%%%%%%%%%%%%%%%%%%%%%%%%%%

\begin{table}[tbp]
\caption{
Coefficients of the perturbation functional $f$ that are collected from the
dimensionless terms of Table I, their theoretical BCS-values in the clean and
dirty superconductors and their experimental estimation in TaN}
\begin{tabular}{llll}
$c_{i}$ & clean & dirty & TaN\\ 
\hline
$c_{2}=2\sigma +\mu +\lambda $ & 0.37  & -0.35 &  \\
$c_{4}=-\sigma +\rho ^{\prime }+\rho ^{\prime \prime }+\tau -2\mu -\lambda $
& 0.50  & 0.10 & 0.30 \\
$c_{6}=\upsilon -\rho ^{\prime }-\rho ^{\prime \prime }+\mu -\tau $ & -0.09
& 0.01 & -0.15\\
\end{tabular}
\end{table}

\begin{table}[tbp]
\caption{
%\mbox 
Structural factors $\beta _{n}^{(N)}$ of the $N=1,2$ quanta 
square and triangular vortex
lattices close to $H_{c2}$  and structural factors $\zeta_{n}^{(N)}$ 
of  one- and two-quanta vortices. }
\begin{tabular}{ccccc}
& \multicolumn{2}{c}{$N=1$} & \multicolumn{2}{c}{$N=2$} \\ 
$\qquad $ & $\qquad \Delta \qquad $ & $\qquad \square \qquad $ & $\qquad
\Delta \qquad $ & $\qquad \square \qquad $ \\ \hline
$\beta _{4}^{(N)}$ & 1.16 & 1.18 & 1.34 & 1.43 \\ 
$\beta _{6}^{(N)}$ & 1.42 & 1.50 & 1.95 & 2.32 \\ 
$\zeta _{4}^{(N)}$ & \multicolumn{2}{c}{1.58} & \multicolumn{2}{c}{1.45} \\ 
$\zeta _{6}^{(N)}$ & \multicolumn{2}{c}{2.00} & \multicolumn{2}{c}{1.75} \\ 
$\zeta _{8}^{(N)}$ & \multicolumn{2}{c}{2.34} & \multicolumn{2}{c}{1.99} \\ 
\end{tabular}
\end{table}

%%%%%%%%%%%%%%%%%%%%%%%%%%%%%%%%%%%%%%%%%%%%%%%%%%%

\mediumtext

\begin{table}[tbp]
\caption{\mbox{Characteristic points at $H$-$T$ diagram of superconductor with $\gamma
=\kappa ^{2}-{1 \over 2}$ close to zero.}}

\begin{tabular}{rll}
4th order coefficient in  &  &
- Tricritical point $T_{2}$ where at $c_{6}>0$ the $H_{c2}(T)$ 
\\
 energy expansion $<$ 0 at: & $\gamma <c_{4}t$ & \ \  transition becomes discontinuous
\\
$H_{c}(T)=H_{c2}(T)$ at: & $\gamma =(c_{4}+c_{6})t$ & - Triple point $L_2$
where at $c_{6}>0$ the $H_{c}(T)$, 
\\
  &           & \ \  $H_{c1}^{\ast}(T)$ and $H_{c2}(T)$ transition lines meet
\\
$H_{c1}(T)=H_{c2}(T)$ at:  & $\gamma =(c_{4}+1.22c_{6})t$ & - Auxiliary point
\\
Short range vortex inte- & &  - Point where 2-quanta vortex decay onto 
\\
 raction is repulsive at: & $\gamma >(c_{4}+1.26c_{6})t$ & \ \ two close-lying 1-quanta vortices
\\
 $\sigma _{ns} < 0$  at:
  & $\gamma<(c_{4}+1.55c_{6})t$ & -  NS interface energy becomes negative
\\
$H_{c}(T)=H_{c1}(T)$ at: & $\gamma =(c_{4}+1.75c_{6})t$ & 
- Triple point $L_1$ where at $c_{6}>0$ the $H_{c}(T)$, 
\\
& & \ \ $H_{c1}(T)$ and $H_{c2}^{\ast }(T)$ transition lines
meet
\\
$H_{c1}^{(1)}(T)<H_{c1}^{(2)}(T)$ at: & $\gamma >(c_{4}+1.89c_{6})t$ & - Two
separate one-quanta vortices are more 
\\
& & \ \ stable then one two-quanta vortex
\\
Long range vortex inte- &                  & - Tricritical point $T_{1}$ where at $c_{6}<0$ the  
\\
raction is attractive at:   & $\gamma <(c_{4}+3c_{6})t$ & \ \ $H_{c1}(T)$ transition becomes discontinuous
\end{tabular}
\end{table}

%%%%%%%%%%%%%%%%%%%%%%%%%%%%%%%%%%%%%%%%%%%%%%%%%%%
\newpage

\narrowtext

\begin{figure}[t]
\vspace{0cm}
\hspace{0cm}
\epsfxsize=8cm
\centerline{\epsffile{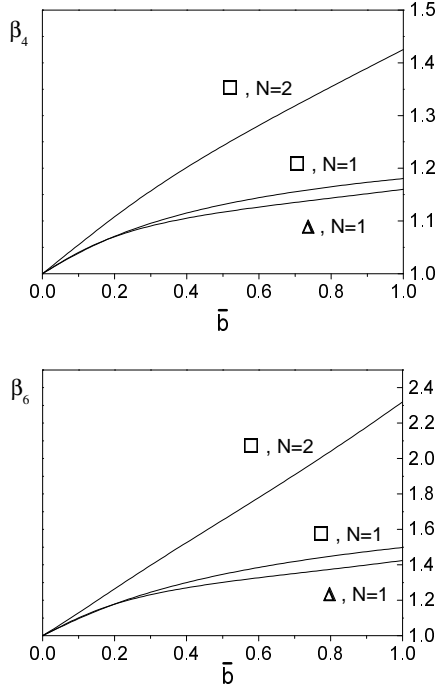}}
\vspace{-0.1cm}

\caption{Structural factors 
$\protect\beta _{4}^{(N)}=<\left| \protect\psi \right|^{4}>
/<\left| \protect\psi \right| ^{2}>^{2}$ and 
$\protect\beta _{6}^{(N)}=<\left| \protect\psi \right| ^{6}>/
<\left| \protect\psi \right|^{2}>^{3}$ 
for triangle and square N-quanta vortex lattices as function of the average
magnetic induction $\bar{b}$.}
\label{Abrfig1}
\end{figure}
%%%%%%%%%%%%%%%%%%%%%%%%%%%%%%%%%%%%%%
%\newpage

%\end{document}

\begin{figure}[t]

\vspace{0cm}
\hspace{0cm}
\epsfxsize=7cm
\centerline{\epsffile{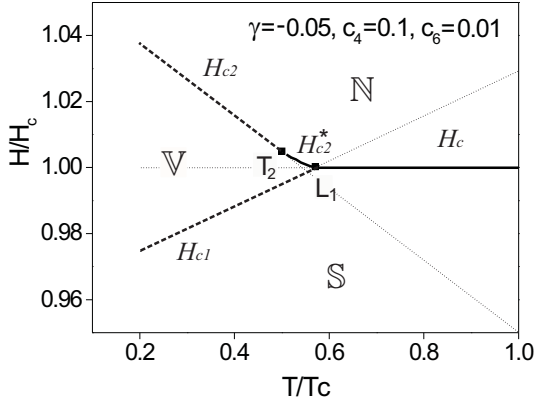}}
\vspace{0cm}

\caption{ $H$-$T$ phase diagram of superconductor with $c_{4},c_{6}>0$ and with $\gamma $
slightly less then zero It includes normal (N), vortex (V) and  Meissner
superconducting (S) phases. Solid and dashed lines corresponds to the discontinuous 
and continuous transitions, dotted lines present the auxiliary critical fields. 
Magnetic field is measured in  units 
of the temperature dependent critical field $H_{c}=H_{c}^{\prime }(T_{c}-T)$.}
\label{Abrfig21}
\end{figure}
%%%%%%%%%%%%%%%%%%%%%%%%%%%%

\begin{figure}[t]
\vspace{0cm}
\hspace{0cm}
\epsfxsize=7cm
\centerline{\epsffile{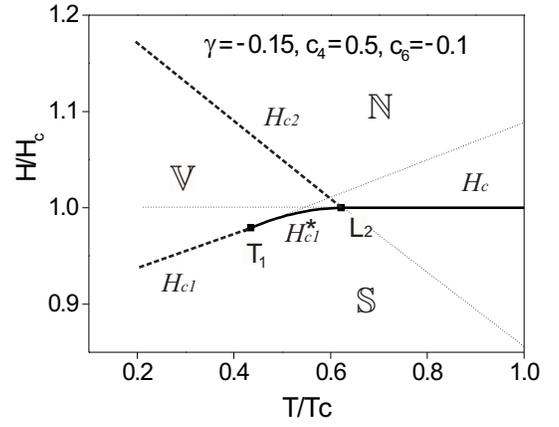}}
\vspace{0cm}
\caption{ The same as Fig. 2 but for $c_{4}>0$, $ 0>c_{6}>-c_{4}/3$ when $\gamma $
is slightly less then zero.  }
\label{Abrfig22}
\end{figure}

%%%%%%%%%%%%%%%%%%%%%%%%%%%%%

\begin{figure}[t]

\vspace{0cm}
\hspace{0cm}
\epsfxsize=7cm
\centerline{\epsffile{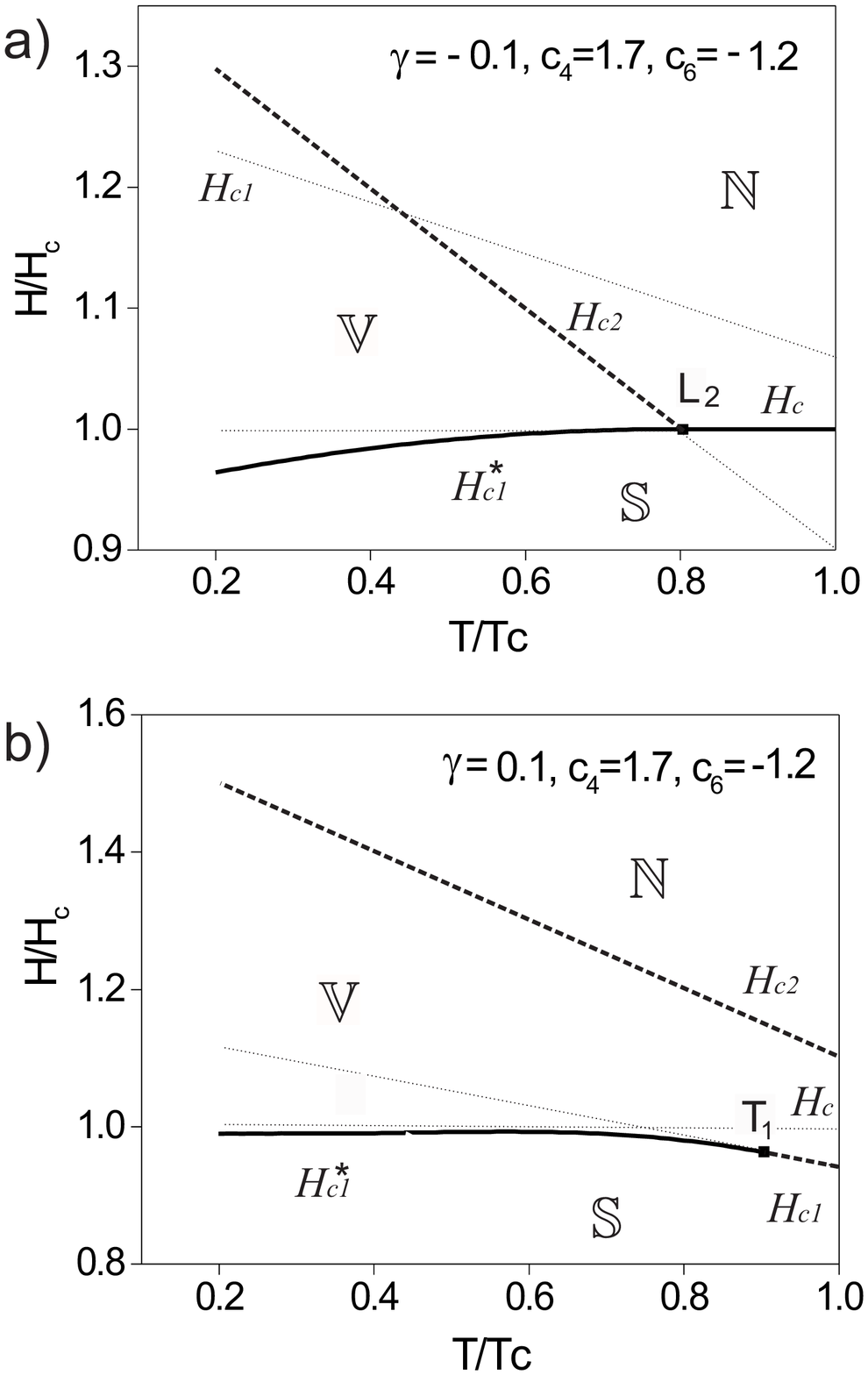}}
\vspace{0cm}

\caption{
The same as Fig. 2 and Fig. 3 but for $c_{4}>0$, $-c_{4}/3>c_{6}$ when
(a) $\gamma $ is slightly less then zero (a) and when  $\gamma $ is
slightly
larger then zero (b). }
\label{Abrfig23}
\end{figure}
\vspace {2cm}

\em

\end{document}